\documentclass[useAMS,usenatbib]{mn2e}
\usepackage{amsmath,fleqn,graphicx,amssymb}
\usepackage{multirow}
\renewcommand{\[}{\begin{equation}}
\renewcommand{\]}{\end{equation}}

\def\ex#1{\left\langle#1\right\rangle}

{\newif\ifnotend
\notendtrue
\def\veclist{ABCDEFGHIJKLMNOPQRSTUVWXYZabcdefghijklmnopqrstuvwxyz.}
\def\top#1#2.{#1}
\def\tail#1#2.{#2.}
\loop\expandafter\xdef\csname v\expandafter\top\veclist\endcsname%
{{\noexpand\bf\expandafter\top\veclist}}
\edef\veclist{\expandafter\tail\veclist}
\if\veclist.\notendfalse\fi\ifnotend\repeat}
\def\d{{\rm d}}

\def\Myr{\,\mathrm{Myr}}
\def\kpc{\,\mathrm{kpc}}

\def\kms{\,\mathrm{km\,s}^{-1}}

\def\mas{\,\mathrm{mas}}

\def\msun{\,{\rm M}_\odot}

\title[The origin of the Gaia phase-plane spiral]
{The origin of the Gaia phase-plane spiral}

\author[James Binney \& Ralph Sch\"onrich]{
  James Binney\thanks{E-mail: binney@physics.ox.ac.uk} and Ralph Sch\"onrich\\  
  Rudolf Peierls Centre for Theoretical Physics, Clarendon Laboratory,
  Oxford, OX1 3PU, UK
}

\begin{document}
\maketitle

\begin{abstract}
A simple model is presented of the formation of the spiral the $(z,v_z)$
phase plane of solar-neighbourhood stars that was recently discovered in Gaia
data. The key is that the frequency $\Omega_z$ at which stars oscillate
vertically depends on angular momentum about the $z$ axis in addition to the
amplitude of the star's vertical oscillations.  Spirals should form in both
$\ex{v_\phi}$ and $\ex{v_R}$ whenever a massive substructure, such as the Sgr
dwarf galaxy, passes through the Galactic plane. The model yields similar
spirals to those observed in both $\ex{v_\phi}$ and $\ex{v_R}$. The primary
driver is the component of the tidal force that lies in the plane. We
investigate the longevity of the spirals and the mass of the substructure,
but the approximations inherent in the model make quantitative results
unreliable. The work relies heavily on a self-consistent, multi-component
model of our Galaxy produced by the AGAMA package for $f(\vJ)$ modelling. 
\end{abstract}


\section{Introduction} \label{sec:intro}

The second data release from the Gaia mission \cite[DR2;][]{BrownDR2} revealed
an unexpected spiral pattern in the phase plane $(z,v_z)$ associated with
oscillations perpendicular to the Galactic plane of stars that lie in a thin
cylindrical shell around the Sun and are located at similar Galactocentric
azimuthal coordinates to that of the Sun \citep{Antoja2018}. The spiral is barely
visible in a plot of the density $\rho$ of stars in the $(z,v_z)$ plane but clearly
visible in a plot in this plane of $\ex{v_\phi}$, the mean value of the component
of velocity in the direction of Galactic rotation. As \cite{Antoja2018}
remark, this feature is suggestive of the phase-winding of a
group of stars that start strongly clumped in the $(z,v_z)$ plane because
such a clump would shear into a spiral of the observed handedness as a
consequence of the vertical period of oscillation being an increasing
function of their amplitudes -- the vertical oscillations of
stars are strongly anharmonic.

While we shall see that this intuition is sound, it does not explain why the
spiral is so much more  clearly visible in $\ex{v_\phi}(z,v_z)$ than in
$\rho(z,v_z)$, nor does it explain the origin of the putative clump.

Here we use the technique of $f(\vJ)$ modelling
\citep{JJB10,PifflPenoyreB,Pascale2018} to show how an `intruder', such
as a dwarf galaxy or a pure dark-matter structure, that passes through the
disc near the pericentre of its orbit generates a feature like that observed.

In Section~\ref{sec:Phi} we outline both the model Galaxy and the data from Gaia
DR2 on which we rely.
Section~\ref{sec:physics} contains the paper's core: a qualitative
explanation of how a passing massive body perturbs the disc leads in
Section~\ref{sec:toy} to a simple
quantitative model that can produce a spiral like that observed. Then in
Section~\ref{sec:move} we
explain how the spiral emerges and investigate its longevity
(Section \ref{sec:longevity}). In Section~\ref{sec:vR} we  show
that a similar but distinct spiral is expected in $v_R$, and we display, for
the first time, this spiral in both the model and Gaia data.
Section~\ref{sec:discuss} discusses the model's strengths and weaknesses, concluding
that full N-body simulation is needed. Section~\ref{sec:conclude} sums up and
looks to the future.

\section{Underpinnings}\label{sec:Phi}
\subsection{The underlying Galaxy model}

Our work involves perturbing a model Galaxy that has been fitted to the
subsample of DR2 that comprises stars with measured line-of-sight velocities
\citep{CropperDR2}. These stars are so numerous and cover such a significant
range of radii that their kinematics when binned in ranges of $R$ and $z$
suffice to constrain strongly the structure of the Galaxy's discs, stellar
halo and dark halo \citep[][in preparation]{BinneyV}. The model is defined by
distribution functions $f(\vJ)$ that are analytic functions of the action
integrals $\vJ=(J_r,J_z,J_\phi)$. The bulge, the stellar halo, the
low-$\alpha$ disc, the high-$\alpha$-disc and the dark halo are all assigned
an $f(\vJ)$ and the parameters within the $f(\vJ)$ are adjusted until the
predicted
stellar kinematics provide reasonable fits to the DR2 data. After each change
of the parameters, the software package AGAMA \citep{AGAMA} is used to
solve for the potential that the  bulge, discs and dark halo jointly
generate alongside a fixed representation of the mass of the interstellar
medium. Hence the final (axisymmetric) model is fully self-consistent, and
strongly constrained within the radial range covered by the DR2 data. It very
nicely reproduces the vertical density profile that \cite{GiRe83} used to
discover the thick disc (now better termed the high-$\alpha$ disc).

\subsection{Data from Gaia DR2}

We use the radial velocity subset of Gaia DR2 \citep{CropperDR2}, which
comprises more than $7$ million stars with full $6$D phase space information,
i.e., positions, parallaxes, proper motions and line-of-sight velocity
measurements. Heliocentric stellar positions and velocities are translated
into the Galactic frame assuming the Sun's Galactocentric radius and vertical
position are $(R_0, z_0) =(8.27, 0.02) \kpc)$ \citep{SchoenrichBD2010,
Schoenrich2012, Joshi2007},
and its Galactocentric motion is $(U_0,V_0,W_0) = (11.1, 250, 7.24)\kms$, so
the Sun is moving inwards and upwards.

We assign distances to stars using the iterative method of
\cite{SchoenrichAumer2017}, which has been validated on Gaia DR1
\citep{Lindegren2016} matched with the RAVE \citep{Kunder2017} and LAMOST
surveys. By using a self-generated prior on distance, this method ensures
unbiased distances for every sample, and statistically validates the
distances. 

The  Gaia DR2 parallaxes have significant zero-point errors
\cite{Lindegren2018}. For the sake of simplicity, we here apply a uniform
parallax offset, $\delta\varpi = -0.048 \mas$ (Sch\"onrich et al.\ in prep).
We use only stars that (i) satisfy the parallax quality cut $\sigma_\varpi /
\varpi < 0.2$, and (ii) have $\ge5$ visibility periods, and (iii) have `excess
noise' $<1 \mas$.  In the plots below, we use stars that lie in the region
$|R-R_0|<0.5\kpc$, $|y|<4\kpc$, $|z|<1.5\kpc$.

\section{The physics of the phase-plane spiral}\label{sec:physics}

From the fact that the spiral appears clearly in a plot of $\ex{v_\phi}$ it
follows that it is associated with a correlation between the in-plane
oscillations (radial epicyclic motion) and oscillations perpendicular to the
plane. The passage of an intruder through the plane will inevitably generate
such correlated oscillations. Indeed, consider an intruder that
passes down through the plane such that its velocity at $z=0$ is in the $-\ve_z$
direction.  Throughout its passage the intruder attracts stars   towards it with the
consequence that the $v_R$  and $v_\phi$ components of each star's velocity
change steadily during the passage. The $v_z$ component, by contrast is
incremented as the intruder approaches from above, but then decremented as
the intruder recedes below the plane. The decrement would cancel the
increment
perfectly if the star in question were stationary during  the passage, but on
account of the star's motion they do not quite cancel, and the star receives
a net kick, which is on average downwards.

\begin{figure}
\includegraphics[width=.9\hsize]{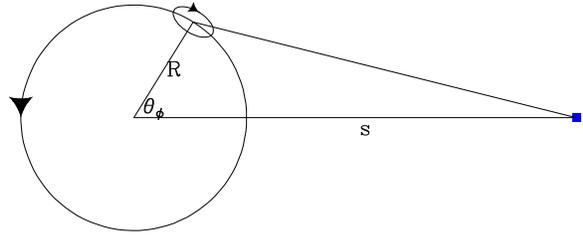}
\caption{Schematic of the impact on a star of an intruder passing through the plane
distance $s$ from the GC.}\label{fig:one}
\end{figure}

We decompose the motion of each star into two parts. The first part is the
star's unperturbed motion, being part circulation and part radial and
vertical oscillation. Averaged over all stars this unperturbed motion brings
stars neither closer to the intruder nor further from it during the passage,
so this motion doesn't give rise to a net vertical impulse on the Galaxy by
the intruder. On the other hand, stars initially at 5 o'clock in
Fig.~\ref{fig:one}, which are being carried towards the intruder by Galactic
rotation, will be closer to the intruder as it recedes than they were as it
approached. Hence these stars will receive a net downward impulse.
Conversely, stars initially at 2 o'clock will receive a net upward impulse.
From this discussion it follows that to lowest order of Fourier expansion in
the azimuth $\phi$ of a star at the instant of pericentre, the vertical
impulse
arising from rotation will be $\propto\sin(\phi-\phi_{\rm intruder})$.

The second part of the motion of stars is the motion arising from the pull of
the intruder. On average this moves stars towards the intruder, and thus
causes the downward pull as the intruder recedes to exceed the upward pull
as the intruder approaches. Hence the pull of the intruder towards its line
of flight always generates a net downward kick. This is essentially the
physics of dynamical friction. 

\begin{figure}
\includegraphics[width=.9\hsize]{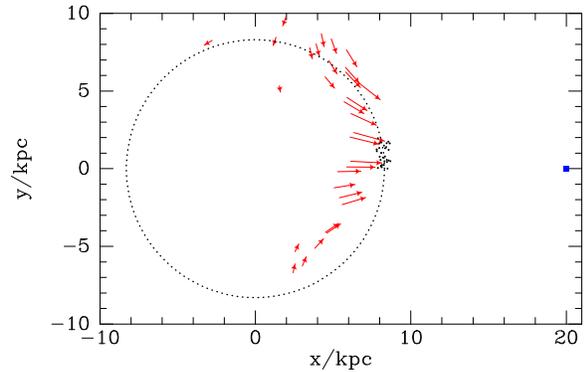}
\caption{The centres of red arrows mark the locations $200\Myr$ ago of stars
that are currently close to the Sun (black dots). The arrows are proportional
to the tidal impulses the stars receive during the passage of the intruder,
which we see crossing the plane $200\Myr$ ago (blue square). The solar circle
is traced by the dotted black line.
}\label{fig:tide}
\end{figure}

\subsection{A toy model}\label{sec:toy}

With AGAMA one can quickly sample the stellar DF subject to any
selection function $g(\vx)$. We used this facility to choose phase-space
coordinates for $N=400\,000$ stars that lie within the region
\[
R_0-0.5\kpc<R<R_0+0.5\kpc\qquad |\phi-\phi_\odot|<0.25\,\hbox{rad}.
\] 
 The orbits of these stars were then integrated backwards for a time $\tau$.
Then we computed the difference between the gravitational acceleration towards
an intruder at the location $(R,z=0,\phi=0)$ on (i) a star located at
$\vx(-\tau)$ and (ii) a star located at the Galactic centre $(R=0,z=0)$. By
subtracting the latter acceleration from the pull of the intruder, we take
cognisance that we are working in the non-inertial frame in which the Galactic
centre remains stationary. The kick stars experience is proportional to the
product $M\times T$ of the intruder's mass and the effective duration of the
passage
\[
T={2p\over v_{\rm intruder}},
\]
 where $p=10\kpc$ is the impact parameter from the perspective of the Solar
neighbourhood and $v_{\rm intruder}\simeq300\kms$ is the intruder's speed.
$M=2\times10^{10}\msun$ and $T=66\Myr$ were adopted for the figures.

Fig.~\ref{fig:tide} shows in black the current
locations of 30 stars from the solar-neighbourhood sample.  The red arrows
are centred on the locations of these stars $\tau=200\Myr$ in the past as the
intruder (blue square) passed through the plane. The lengths of the arrows
are proportional to the kicks $\delta\vv$ that these stars had just
received from the intruder after allowing for the fact that the reference
frame has been kicked as if it were a particle at the Galactic centre. 

The acceleration computed above multiplied by a characteristic timescale of
the passage yields a velocity change $\delta\vv_\parallel$ that lies within the
plane. To this we must add a (downward) component $\delta v_\perp$ that
arises because stars move during the encounter. As explained at the head of
this section, the Fourier expansion of this velocity change in the azimuths of
stars will be dominated by a constant term associated with standard dynamical
friction and a term proportional to $\sin(\phi-\phi_{\rm intruder})$. The
magnitude of both terms will be scale with 
the modulus of the in-plane component. Pending a more elaborate treatment, we
assume both terms have the same magnitude and adopt
\[\label{eq:down}
\delta v_\perp=0.4|\delta v_\parallel|\,[1-\sin(\phi-\phi_{\rm intruder})].
\]
 The minus sign ensures that the downward kick is larger for stars that are
initially moving towards the intruder and the number $0.4$ is required to
yield satisfactory plots.  Eliminating the $\phi$-dependent
factor above made no significant difference to the results.

In summary, except when estimating the vertical kick $\delta v_\perp$, we
imagine that the intruder's gravitational field is a Dirac delta-function in
time with a magnitude that is proportional to the intruder's mass times the
approximate duration of the passage in the real world. Thus we are
essentially working in the impulse approximation \citep[\S7.2.1][]{GDI}.

\begin{figure}
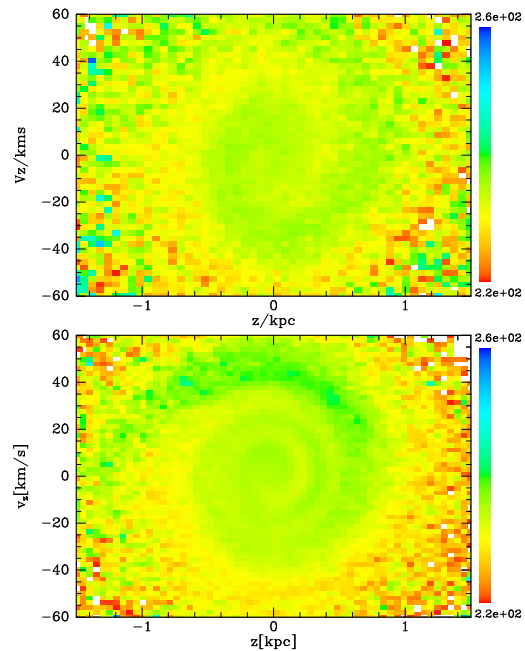

\centerline{\includegraphics[width=.8\hsize]{figs/plot_spiral2.ps}}
\centerline{\includegraphics[width=.8\hsize]{figs/rvsR0Vp.ps}}
\caption{Upper panel: the average value of $v_\phi$ within the $(z,v_z)$ phase plane
computed for an impulsive passage of an intruder $200\Myr$ ago at a
Galactocentric distance of $20\kpc$. Lower panel: the same average for stars
in DR2.}\label{fig:plot_spiral}.
\end{figure}

From the velocity $\vv(-\tau)$ of each star that we  have computed by
backwards integration we subtract $\delta\vv$ computed as above to obtain the
star's velocity before the intruder appeared. At the phase-space
location $[\vx(-\tau),\vv(-\tau)-\delta\vv]$ we evaluate the stellar DF
$f(\vJ)$ (the sum of the DFs of bulge, stellar halo and discs) by using the
St\"ackel Fudge \citep{JJB12:Stackel} to evaluate the actions. By Liouville's
theorem, this value $f_0$
is the actual phase-space density at the current location of the star near
the Sun. By contrast, the sampling density of our sample is $f_{\rm
s}=f(\vJ)$, where $\vJ$ comprises the actions of the sampling point and of the orbit
computed back to $\vx(-\tau)$. By the principle of Monte-Carlo integration,
a valid estimate of the current expectation value of any phase-space variable $q$ is
\[
\ex{q}={1\over N}\sum_{i=1}^N q_i{f_{0i}\over f_{{\rm s}i}}.
\]
 The result of applying this formula to $q=v_\phi$ is shown in the upper
panel of Fig.~\ref{fig:plot_spiral}. A spiral is evident that is remarkably
similar to that seen in the lower panel, which after \cite{Antoja2018}
shows $\ex{v_\phi}$ in the DR2 data.

Fig.~\ref{fig:dens}  shows the current density of model stars in the
$(z,v_z)$ plane: no spiral is evident just as in the corresponding plot in
\cite{Antoja2018}.

\begin{figure}
\centerline{\includegraphics[width=.8\hsize]{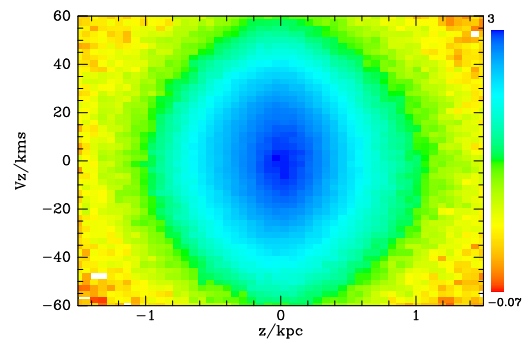}}
\caption{Density of model stars in the $(z,v_z)$ plane.}
\end{figure}\label{fig:dens}

\begin{figure}
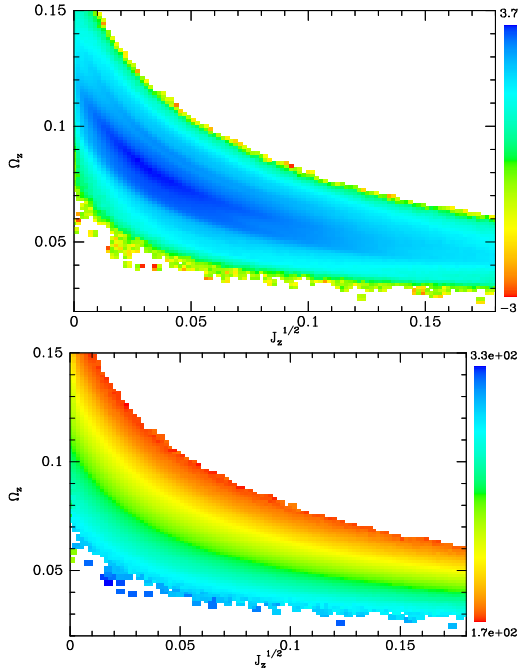

\includegraphics[width=.8\hsize]{figs/findJ6.ps}
\includegraphics[width=.8\hsize]{figs/findJ7.ps}
\caption{Upper panel: from Gaia DR2 we show the density of stars in
the plane $(J_z,\Omega_z)$. Lower panel: the  mean value of $v_\phi$ within this
plane.}\label{fig:findJ}
\end{figure}

\subsection{Moving around your ellipse}\label{sec:move}

\begin{figure}
\centerline{\includegraphics[width=.6\hsize]{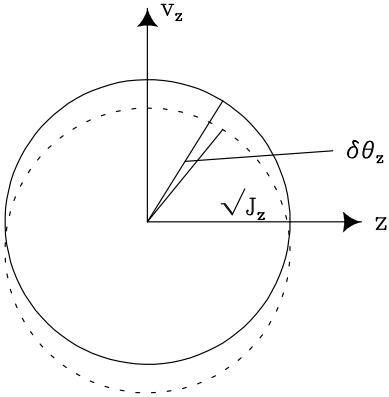}}
\caption{Schematic of the $(z,v_z)$ plane showing the effect on the DF of
decrementing all values of $v_z$. Initially the full ellipse, a curve of
constant $J_z$, is a contours of $f$. After velocities have been decremented,
this contour becomes the dashed curve, which is not a curve of constant
$J_z$. Consequently,  on curves of constant $J_z$ stars are concentrated towards
$\theta_z=\pi$ since the origin of $\theta_z$ is conventionally at the top of
the curves. }\label{fig:shift}
\end{figure}

To obtain a better understanding of how the phase-plane spiral forms,
the upper panel of Fig.~\ref{fig:findJ} shows the distribution of stars in the
plane $(\surd J_z,\Omega_z)$. $\surd J_z$ is the natural radial coordinate in
the $(z,v_z)$ plane: in this plane stars move clockwise on ellipses of area
$2\pi J_z$. Stars lie in a broad swath that extends from large $\Omega_z$ at
small $J_z$ to small $\Omega_z$ at large $J_z$, and it was to this trend of
$\Omega_z$ with $J_z$ that \cite{Antoja2018} appealed in their picture of
phase-wrapping of a clump of stars. As remarked in the Introduction, the
problem with this explanation is the invisibility of a spiral in the density
$\rho(z,v_z)$. The lower panel of Fig.~\ref{fig:findJ} reveals the reason for
the width of the swath by plotting $\ex{v_\phi}$, the average value of
$v_\phi$ in each pixel. This forms a perfect rainbow, with low $\ex{v_\phi}$
at the top of the swath and high $\ex{v_\phi}$ at the bottom. This
correlation arises because all stellar frequencies decrease as one moves out through
the Galaxy, and in the solar neighbourhood $v_\phi$ is tightly connected to
$J_\phi$, which controls an orbit's guiding-centre radius.

By Jeans theorem, in a relaxed Galaxy the distribution of stars is stratified
in the $(z,v_z)$ plane on ellipses of constant $J_z$, so it is independent of
$\theta_z$.  The intruder shifts
the distribution of stars down in the $(z,v_z)$ plane so it is no longer
stratified by $J_z$ (Fig.~\ref{fig:shift}). Hence the distribution is now a
function of the conjugate angle $\theta_z$ in addition to $J_z$, with a
concentration of stars at $\theta_z=\pi$.\footnote{We adopt the convention of
{\tt TM} \citep{JJBPJM16} that $\theta_z=0$ as stars move up through the
plane.} After the perturbation has been applied, each star moves around its
newly assigned ellipse at a rate $\Omega_z$ that varies systematically with
$v_\phi$. Hence in the top panel of Fig.~\ref{fig:plot_spiral} the colour of
an ellipse varies with location $\theta_z$ around it. On account of the
dependence of $\Omega_z$ on $J_z$, the colouration evolves on smaller
ellipses faster than on larger ellipses.  Hence the emergence of the spiral
depends essentially on both functional dependencies of
$\Omega_z(J_\phi,J_r)$.

\subsection{How long does a spiral last?}\label{sec:longevity}

\begin{figure}
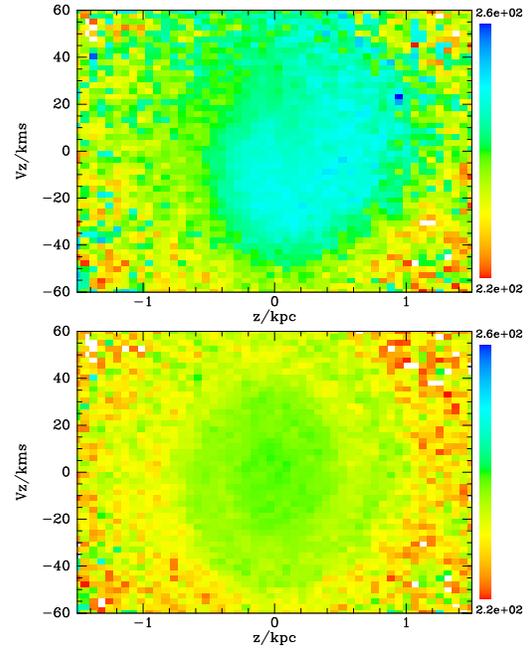

\centerline{\includegraphics[width=.8\hsize]{figs/plot_spiral1.ps}}
\centerline{\includegraphics[width=.8\hsize]{figs/plot_spiral3.ps}}
\caption{Upper panel: $\ex{v_\phi}$ $100\Myr$ after a passage. Lower panel:
$\ex{v_\phi}$ $400\Myr$ after a passage.}\label{fig:other_t}
\end{figure}

In Fig.~\ref{fig:other_t} we show $\ex{v_\phi}$ in the phase plane $100\Myr$
(upper panel) and $400\Myr$ (lower panel) after an intruder passed through the plane. In
each case the intruder passed through the disc at what was at that time the
mean azimuth of the sampled stars. Hence the stars that contribute to the
upper panel have gone half way round the Galaxy since they were kicked, while
those in the lower panel have been round twice. In the upper panel the spiral
is still under-developed, while in the lower panel it has been would up so
much that with $400\,000$ stars its form is becoming unclear. Thus the observed
spiral would seem to have been created $200^{+200}_{-50}\Myr$ ago.

\subsection{A spiral in $v_R$}\label{sec:vR}

\begin{figure}
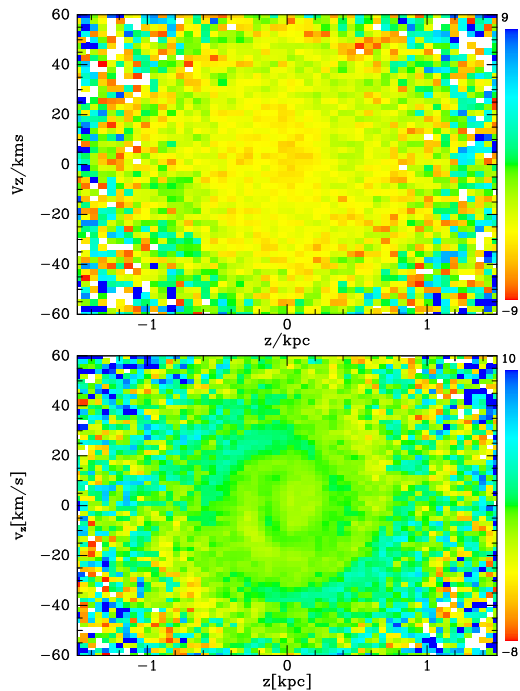

\centerline{\includegraphics[width=.8\hsize]{figs/plot_spiral5.ps}}
\centerline{\includegraphics[width=.8\hsize]{figs/rvs8_3VR.ps}}
\caption{Upper panel: $\ex{v_R}$ in the phase plane at $200\Myr$. Lower
panel: the same mean over stars in DR2 that have $|v_R|<30\kms$.}\label{fig:vR}
\end{figure}

The intruder pulls stars outwards and thus causes an overdensity of stars in
the $(R,v_R)$ phase plane around $\theta_R=\pi/2$ (maximum outward speed) --
the phase-plane geometry is extremely similar to that shown in
Fig.~\ref{fig:shift}.  If at this instant we were to plot $\ex{v_R}$ in the
$(z,v_z)$ plane, the plot would show a coherent region of positivity rather
than the uniform zero characteristic of an equilibrium disc. Now focus on the
stars that lie at a point of the $(z,v_z)$ plane where $\ex{v_R}>0$. They
have a common value of $J_z$ and move around a common ellipse, but at rates
$\Omega_z$ that increase as $J_\phi$ decreases. As $J_\phi$ decreases
$\Omega_r$, like $\Omega_z$, increases, so the $\theta_r$ values of the stars
that move fastest around the ellipse increase fastest. As $\theta_r$ changes,
so do the $v_R$ values of these stars, so the contribution of these stars to
$\ex{v_R}$ changes faster than do the contributions of the groups of stars
that lag them. It follows that the blob of colour signalling $\ex{v_R}>0$ is
sheared into a spiral that differs in the tightness of its winding from the
spiral seen in $\ex{v_\phi}$ because $v_\phi$ and $v_R$ evolve differently as
$\theta_r$ increments.

The upper panel of Fig.~\ref{fig:vR} shows $\ex{v_R}$ in the phase plane
$200\Myr$ after a passage. As expected, a spiral is evident that is similar
to, but different from, that seen in the corresponding plot for
$\ex{v_\phi}$. The lower panel of Fig.~\ref{fig:vR} shows a similar spiral in
the DR2 data. The corresponding plot in \cite{Antoja2018} does not show a
spiral: the key to unveiling a spiral in $\ex{v_R}$ is to restrict the sample
to thin-disc stars by imposing the selection $|v_R|<30\kms$.

Plots of $\ex{v_R}$ at $100$ and $400\Myr$ do not
manifest spirals although they do show evidence for disturbance: $\ex{v_R}$ varies
significantly around ellipses of constant $J_z$.

\section{Discussion}\label{sec:discuss}

We have supposed the action of the intruder to be impulsive in order to
obtain the simplest physical picture that can explain the observed
phenomenon. In reality the impulse approximation will provide only a mediocre
quantitative account of the passage through pericentre of a galaxy such as
Sgr. Moreover, in the limit $v_{\rm intruder}\to\infty$ in which the impulse
approximation becomes exact, the phenomenon under discussion will disappear
because in this limit $\delta v_\perp/\delta v_\parallel\to0$. Hence
quantitative results extracted from the model presented here should be regarded
with some scepticism.

However, the major weakness of the present model is not its use of the
impulse approximation, which has has a history of being more successful than
one has a right to expect \citep[][\S7.2.1]{Alladin1982, GDI}, but in the use
of unperturbed frequencies $\Omega_z$: when a whole section of the disc is
pulled down, stars will oscillate vertically with significantly longer
periods than $2\pi/\Omega_z$ \citep{HunterToomre}. What the present model
gets right is the insight that the frequencies of vertical and radial oscillations
decrease as $J_\phi$ and thus $v_\phi$ increase. Qualitatively, the
explanation we have given is sound even though it is quantitatively
significantly off.

Careful comparison of the upper and lower panels of
Figs.~\ref{fig:plot_spiral} and \ref{fig:vR} reveals quantitative weaknesses.
Most obviously, the model spirals seem stretched along the velocity axis with
respect to those in the DR2 data. The model spirals are possibly also less
tightly wound that the data spirals, although this conclusion is sensitive to
one's interpretation of faint features near the centres of the spirals.

The mass and impact parameter of the intruder are significant issues. Uncomfortably
large values of the product $MT$ of intruder mass and passage duration were
required to obtain clear model spirals. With such large values of $MT$ stars
receive significant kicks and the ratio $f_0/f_{\rm s}$ of the values taken
by the unperturbed DF at a star's location before and after kicking differ
considerably from unity.  Comparison of the colouring in the upper and lower
panels of Fig.~\ref{fig:vR} suggests that the model passages are generating
larger values of $v_R$ than they should be.

Another issue is the adopted magnitude of the applied downward kicks $\delta
v_\perp$
(eqn.~\ref{eq:down}). As was explained at the start of
Section~\ref{sec:physics}, the
relationship of this kick to the in-plane kick depends on the intruder's
speed. We have not attempted to compute it, but simply adopted a value that
generates reasonable spirals. The value we have adopted is on the  large
side, but smaller values don't yield good figures.

Large values of $\delta v_\perp$ imply a slow intruder, which implies
pericentre of a tightly-bound, low eccentricity orbit. The Sgr dwarf has just
such an orbit \citep{DierickxLoeb2017}. However, our impulse approximation is
least justifiable in this regime: an individual passage will be slow, and the
data may contain signals from more than one passage, so our picture could be
seriously over-simplified. 

To overcome the weakness of our approach both with respect to the likely
slowness and recurrence of the passages and the need to recognise that whole
sections of the disc are moved up and down, full N-body modelling is
required. Such models are orders of magnitude more expensive than our
approach, and they don't provide the same insight, but they are essential for
proper exploitation of the beautiful data that are now to hand.

\section{Conclusions}\label{sec:conclude}

A simple model based on the impulse-approximation in which an intruder plucks
the disc out and down produces a spiral in the $(z,v_z)$ plane very like that
discovered in $\ex{v_\phi}$ by \cite{Antoja2018}. We have shown that a
similar spiral in $\ex{v_R}$ can be teased from the same data, and that our
model can reproduce this spiral also. There are indications that the intruder
is not on a highly eccentric orbit, with the consequence that our use of the
impulse approximation is questionable and that full N-body modelling will be
required to explain the data satisfactorily.

Whatever the orbit of the intruder, our model necessarily neglects the fact
that the intruder plucks downwards whole segments of the disc rather than individual
stars from the disc, with the consequence that our model uses vertical
frequencies that are systematically too high. This is a serious weakness of
the model, and relegates it to `toy' status: it captures important aspects of
the problem and delivers valuable insights but over-simplifies to the point
that it cannot be trusted quantitatively.

This investigation illustrates nicely an aspect of the power latent in
$f(\vJ)$ modelling.  Specifically it shows the value of being able to sample
our corner of the Galaxy densely: if we assume the disc has scalelength
$R_\d=2.5\kpc$, only $0.38\%$ of the disc's stars lie in the region we have
sampled, so an N-body model providing the same resolution in the $(z,v_z)$
plane would require in excess of 100 million disc stars, not counting
particles needed to represent the bulge, dark halo, etc. A simulation on this scale is
computationally expensive. While it is true that with current $f(\vJ)$
software it is not possible to model correctly the collective response of the
disc, so ultimately N-body models are essential, the software does provide the initial
data for an N-body simulation that most closely approach an equilibrium and
thus eliminate unwanted transients \citep{AGAMA}. Consequently, the way
forward would seem to be speedy exploration of possibilities by $f(\vJ)$
modelling followed by a few high-quality N-body simulations with initial
conditions furnished by the most successful $f(\vJ)$ models.


\section*{Acknowledgements}

This work was stimulated by conversations with J Bland-Hawthorn.

\def\physrep{Phys.~Reps}
\bibliographystyle{mn2e} \bibliography{/u/tex/papers/new_refs}
\end{document}